\documentclass[apj]{emulateapj}

\slugcomment{Last updated: \today}
\shorttitle{Evolution of Galaxy Morphology} 
\shortauthors{Hwang \& Park}

\begin{document}

\title{Evidence for Morphology and Luminosity Transformation of Galaxies at High Redshifts}

\author{Ho Seong Hwang and Changbom Park}
\affil{School of Physics, Korea Institute for Advanced Study, Seoul 130-722, Korea}
\email{hshwang@kias.re.kr, cbp@kias.re.kr}

\begin{abstract}

We study the galaxy morphology-luminosity-environmental relation
  and its redshift evolution using a spectroscopic sample of galaxies 
  in the Great Observatories Origins Deep Survey (GOODS).
In the redshift range of $0.4\leq z\leq1.0$
  we detect conformity in morphology between neighboring galaxies.
The realm of conformity is confined within the virialized region associated with
  each galaxy plus dark matter halo system.
When a galaxy is located within the virial radius of its nearest neighbor galaxy, 
  its morphology strongly depends on the neighbor's distance and morphology: 
  the probability for a galaxy to be an early type ($f_E$) strongly
  increases as it approaches an early-type neighbor, but
  tends to decrease as it approaches a late-type neighbor.
We find that $f_E$ evolves much faster
  in high density regions than in low density regions, and that
  the morphology-density relation becomes significantly weaker at $z\approx 1$.
This may be because
  the rate of galaxy-galaxy interactions is higher in high density regions,
  and a series of interactions and mergers over the course of galaxy life
  eventually transform late types into early types.
We find more isolated galaxies are more luminous, which supports
luminosity transformation through mergers at these redshifts.
Our results are consistent with those from nearby galaxies, 
  and demonstrate that galaxy-galaxy interactions have been strongly affecting
  the galaxy evolution over a long period of time.
\end{abstract}



\section{Introduction}
The role of environment in determining galaxy properties
  is one of key issues in galaxy formation and evolution.
In particular, galaxy morphology is known to depend on environment.
It was first noted by \citet{hh31} who found 
  a large population of ellipticals and lenticulars in galaxy clusters.
Later, systematic studies for the connection between galaxy morphology
  and environment suggested the morphology-radius relation \citep{oem74}
  and the morphology-density relation (MDR; \citealt{dre80}).
MDR was detected in the group environment \citep{pg84},
  and was also found in galaxy clusters at redshifts up to 1 \citep{dre97,treu03,smi05,post05}.

Since the Sloan Digital Sky Survey (SDSS; \citealt{york00})
  and Two Degree Field Galaxy Redshift Survey (2dFGRS; \citealt{col01})
  have produced unprecedentedly large photometric and spectroscopic data of nearby galaxies,
  the environmental dependence of galaxy properties in local universe has been extensively revisited
  (e.g., \citealt{goto03,bal04a,bal04b,tan04,bla05,wei06,park07,park08,pc09,ph08}).
Among them, \citet{park07} found that the environmental dependence of various galaxy properties
  is almost entirely due to their correlation with morphology and luminosity that depend on the local density. 
If both morphology and luminosity are fixed, galaxy properties such as color, color gradient,
  concentration, size, velocity dispersion, and star formation rate (SFR), are nearly independent of
  the local density.
This study was extended by \citet{park08} who took into account the effect
  of the nearest neighbor galaxy.
They found that galaxy morphology depends critically on the small-scale environment,
  which is characterized by the morphology of the nearest neighbor galaxy 
   and the mass density due to the nearest neighbor galaxy,
  in addition to the luminosity and the large-scale density.
They suggested a unified scenario that 
  the morphology and luminosity of a galaxy change 
  through a series of galaxy-galaxy interactions and mergers.

More recently, \citet{pc09} investigated the dependence of galaxy properties 
  on both the small- and large-scale environments.
They found two characteristic pair-separation scales where the galaxy
  properties abruptly change: 
  the virial radius $r_{\rm vir,nei}$ of the nearest neighbor galaxy 
  where the effects of galaxy interaction emerge and
  $\sim$0.05 $r_{\rm vir,nei}$ where the galaxies in a pair start to merge.
The role of large-scale density is weak when morphology and luminosity are fixed.
\citet{ph08} reported that the morphology transformation in massive
  galaxy clusters is driven by hydrodynamic interactions between galaxies
  and by repeated gravitational interactions among galaxies or 
  between galaxies and their host cluster.
Surprisingly, the galaxy morphology does not depend on
  the local galaxy number density at fixed luminosity and fixed nearest neighbor separation.
They found that the morphology-radius relation exists within the cluster virial radius
  but that galaxy morphology is determined almost entirely by the nearest neighbor distance
  and morphology outside the cluster virial radius.
The MDR is only an apparent phenomenon through the statistical correlation
  of the local galaxy number density with luminosity and the nearest neighbor distance.

The MDR beyond the local universe, was investigated mostly 
  in high density regions of galaxy clusters
  (\citealt{dre97,treu03,smi05,post05}; see also \citealt{pog08}).
However, thanks to the recent large, deep-field surveys such as 
  the Great Observatories Origins Deep Survey (GOODS; \citealt{gia04}),
  the All-Wavelength Extended Groth Strip International Survey (AEGIS; \citealt{dav07}),
  the VIMOS VLT Deep Survey (VVDS; \citealt{lef05}),
  the Cosmic Evolution Survey (COSMOS; \citealt{sco07}), and
  the Canada-France-Hawaii Telescope Legacy Survey\footnote{http://www.cfht.hawaii.edu/Science/CFHLS/} 
  (CFHTLS),
  environmental dependence of high redshift `field' galaxy properties has started to be explored:
  MDR \citep{nui05,cap07,wel07}, color-density relation (CDR) \citep{coo07,cuc06,cas07}, 
  and SFR-density relation \citep{elb07,coo08}.
The MDR and CDR observed locally are also found at high redshifts ($z\sim1$),
  but the slope of the relation appears to be different from the local one.
It means that galaxy morphology or color evolves differently
  depending on the background density,
  in the sense that the stronger evolution is seen in high density regions.
Interestingly, the SFR-density relation at high redshifts ($z\sim1$)
  is reversed compared to the local one.
Namely, the SFR of high redshift galaxies
  is larger in high density regions than in low density regions \citep{elb07,coo08}.

These observational findings raised important questions:
  (1) whether the correlations between the galaxy properties
  and the local density are the consequence of environmental-driven evolution;
  (2) which relation is the most fundamental one \citep{coo06,coo07,pog08}.
The physical mechanism of environmental effects on 
  the galaxy properties is also poorly understood \citep{coo06}.
Moreover, there are few studies that focus on the role of the nearest neighbor galaxy,
  which turned out to be very important 
  in the evolution of galaxy morphology and luminosity \citep{park08}.
Since the galaxy interactions and mergers with neighbor galaxies 
  may be more frequent at high redshifts,
  it is necessary to investigate the role of neighbor galaxies
  in determining the properties of high redshift galaxies
  hoping that the main mechanism for galaxy evolution be understood.

In this paper,
  we study the morphology and luminosity of 
  high redshift galaxies adopting the method similar to that used for nearby SDSS galaxies \citep{park08}.
Section \ref{data} describes the observational data used in this study.
Environmental dependence of galaxy morphology and luminosity 
  is given in \S \ref{results}.
Discussion and summary are given in \S \ref{discuss} and \S \ref{sum}, respectively.
Throughout this paper we adopt a flat $\Lambda$CDM cosmological model with density parameters 
$\Omega_{\Lambda}=0.73$ and $\Omega_{m}=0.27$.

\section{Data}\label{data}
\subsection {Observational Data Set}\label{obsdata}

We used a spectroscopic sample of galaxies in GOODS.
GOODS is a deep multiwavelength survey covering two carefully selected regions
  including the Hubble Deep Field North (HDF-N) and the Chandra Deep Field South (CDF-S).
Hereafter, two GOODS fields centered on HDF-N and CDF-S are called
  GOODS-North and GOODS-South, respectively.
Total observing area is approximately 300 arcmin$^2$ and
  each region was observed by NASA's Great Observatories ({\it HST}, {\it Spitzer} and {\it Chandra}),
  ESA's {\it XMM-Newton}, and several ground-based facilities.
{\it HST} observations with Advanced Camera for Surveys (ACS) were conducted
  in four bands: $B$ (F435W, 7200s), $V$ (F606W, 5000s), $i$ (F775W, 5000s), and $z$ (F850LP, 10,660s).
The drizzled images have a pixel scale of $0.03\arcsec$ pixel$^{-1}$ 
  and point-spread function FWHM of $\sim$0.1$\arcsec$.
Spectroscopic data for GOODS sources are enormous in the literature:
GOODS-North \citep{coh00,cow04,wir04,red06} and 
GOODS-South \citep{szo04,lef04,mig05,van05,van06,van08,rav07,pop09}.
Among the sources in the ACS photometric catalog,
  we used 4443 galaxies whose reliable redshifts are available in the literature
  for further analysis.


\begin{figure}
\plotone{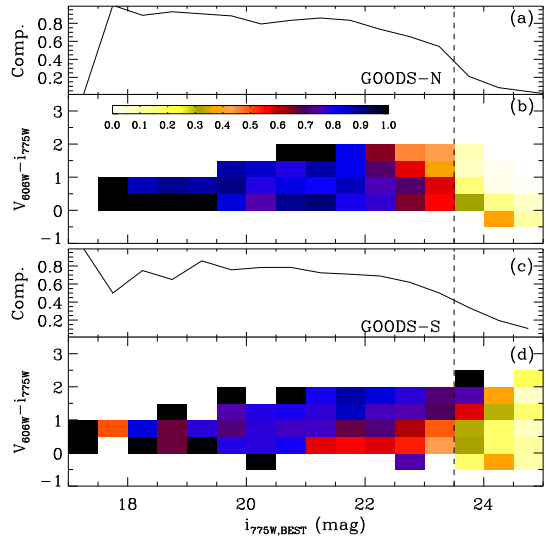}
\centering
\caption{Spectroscopic completeness as a function of $i$-band apparent magnitude ({\it a}) and 
  observed color and apparent magnitude ({\it b}) of galaxies in GOODS-North. 
Those for GOODS-South are in ({\it c}) and ({\it d}). 
Vertical dashed lines indicate the apparent magnitude limits used in this study.
}\label{fig-comp}
\end{figure}

Since we are going to investigate the effects of the nearest neighbor galaxies,
 it is important to identify genuine neighbor galaxies.
In addition, 
  since we use a spectroscopic catalog of galaxies combined from various redshift surveys
  with diverse selection criteria, our sample is heterogeneous.
Therefore, it is necessary to know the completeness of our spectroscopic sample accurately.
To compute the spectroscopic completeness,
  we used a sample from the ACS photometric catalog with the objects having 
  SExtractor stellarity class greater than $0.79$.
The value was chosen from the stellarity distribution of genuine stars 
  confirmed by the spectroscopic observation.
In Figure \ref{fig-comp}, 
  we plot the completeness of each survey 
  as a function of the observed magnitude (SExtractor `BEST' magnitude) and color.
It is seen that the completeness decreases significantly near $i_{775W,BEST}\sim23.5$.
The mean completeness at $i_{775W,BEST}\leq23.5$ 
  is 69\% and 63\% for GOODS-North and -South, respectively.
It is noted that the completeness below the magnitude limit
  changes significantly with color,
  which can cause a bias in the studies of galaxy morphology.

\begin{figure}
\plotone{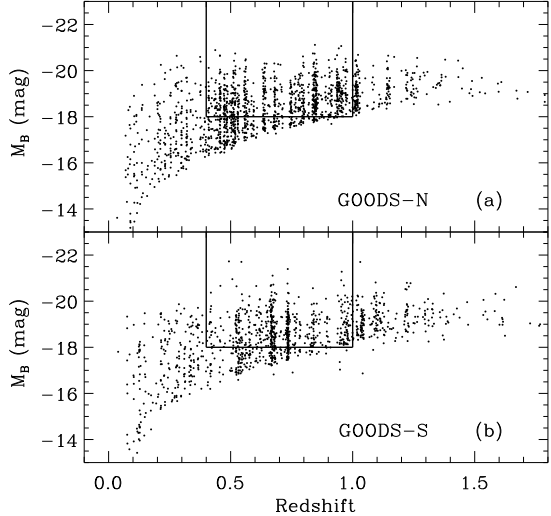}
\centering
\caption{Evolution corrected, rest frame $M_B$  vs. redshift for the spectroscopic sample 
  of galaxies in GOODS-North ({\it a}) and GOODS-South ({\it b}). 
Solid lines define the volume-limited sample used in this study.
}\label{fig-hub}
\end{figure}

The rest frame $B$-band absolute magnitude $M_B$ of galaxies is computed based on the ACS photometry 
  with Galactic reddening corrections \citep{sch98} and $K$-corrections \citep{bla07}.
The evolution correction (an increase of $1.3M_B$ per unit redshift) was applied to 
  compute the final rest frame $M_B$ \citep{fab07}.
In Figure \ref{fig-hub}, 
  we show the evolution corrected, rest frame $M_B$ as a function of redshift for the 
  sample of galaxies with $i_{775W,BEST}\leq23.5$. 
We finally define a volume-limited sample of 1332 galaxies 
  with $M_B\leq-18.0$ and $0.4\leq z\leq1.0$ for further analysis.

\subsection {Morphology Classification}\label{morph}

For the volume-limited sample of galaxies shown in Figure \ref{fig-hub},
  we visually inspected the images in individual ($Bviz$) bands and 
  $Bvi$ pseudo-color images.
We divided the galaxies into two morphological types: early types (E/S0) and late types (S/Irr).
Early-type galaxies are those with little fluctuation in the surface brightness 
  and color and with good symmetry,
  while late-type galaxies show internal structures and/or color variations in the pseudo-color images.
However, the total color itself is not used as a classification criterion.
We checked our results by comparing with the morphological classification of \citet{bun05} 
  who used the same ACS images as ours,
  and found that 98\% of our classifications agree with those of Bundy et al.
Some galaxies that were classified as early types in \citet{bun05}
  are found to be late types in this study
  because of color variations in the pseudo-color images.

To see how often we classify galaxies differently because we classified galaxies
  in different wavelength bands across redshift, we made the following experiment.
We classified galaxies at $0.4<z<0.6$ brighter than $M_B=-18.0$ using their
  $v$- or $i$-band images separately, and checked if the morphological types of each galaxy
  agree with each other.
At $z=0.9$ the $i$-band is centered at $\sim$4000$\AA$ in the rest frame, and this wavelength
  roughly corresponds to the $v$-band at $z\approx0.5$.
We found no galaxy was assigned different morphology.
Therefore, our morphology classification is not expected to be affected by the redshift effects.

\begin{figure}
\plotone{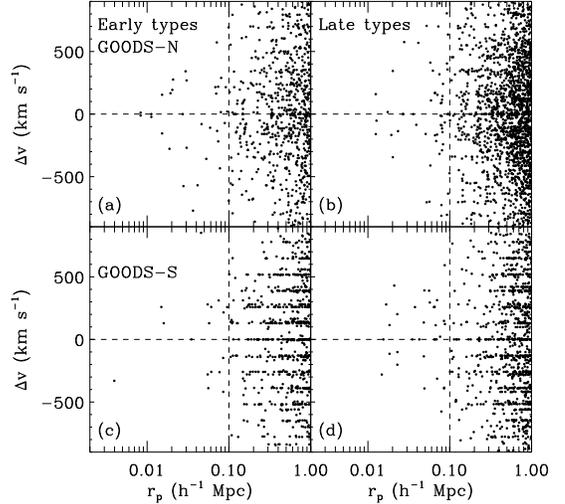}
\centering
\caption{Velocity difference between the target galaxies with $-18.5\geq M_B>-20.0$ and 
  their neighbors brighter than $M_B+0.5$ as a function of the projected separation.
Early- and late-type target galaxies are in left and right panels, respectively.
Galaxies of GOODS-North are in ({\it a}) and ({\it b}), and
  those of GOODS-South in ({\it c}) and ({\it d}).
}\label{fig-veldiff}
\end{figure}

\subsection {Galaxy Environment}\label{environ}

We consider two kinds of environmental factors:
  a surface galaxy number density 
  estimated from five nearest neighbor galaxies ($\Sigma_5$)
  as a large-scale environmental parameter, and
  the distance to the nearest neighbor galaxy ($r_p$)
  as a small-scale environmental parameter.

The background density, $\Sigma_5$, is defined by $\Sigma_5=5(\pi D^2_{p,5})^{-1}$.
$D_{p,5}$ is the projected proper distance to the 5th-nearest neighbor.
The 5th-nearest neighbor of each target galaxy was identified among the neighbor galaxies
  with $M_B\leq-18.0$
  that have velocities relative to the target galaxy less than $1000$ km s$^{-1}$
  to exclude foreground and background galaxies.

To define the small-scale environmental parameter attributed to the nearest neighbor,
  we first find the nearest neighbor of a target galaxy
  that is closest to the target galaxy on the projected sky
  and satisfies the conditions of magnitude and relative velocity.
We searched for the nearest neighbor galaxy among galaxies
  that have magnitudes brighter than $M_B=M_{B,\rm target}+0.5$ and 
       have relative velocities less than $\Delta v=|v_{\rm neighbors}-v_{\rm target}|=600$ km s$^{-1}$
       for early-type target galaxy and less than $\Delta v=400$ km s$^{-1}$ for late-type target galaxy.
These values are the same as those used for selecting the nearest neighbor galaxy in the SDSS data \citep{park08}.
Since we use the volume-limited sample of galaxies with $M_B\leq-18.0$,
  we study only the target galaxies brighter than $M_{B,\rm target}=-18.5$
  so that their neighbors are complete.

In Figure \ref{fig-veldiff}, 
  we plot the velocity difference for all neighbor galaxies 
  with velocity difference less than $900$ km s$^{-1}$.
To fit the velocity distribution,
  we used the Gaussian plus constant model for each type of target galaxies
  by combining the data in two surveys.
We obtained, for the galaxies 
  at projected proper distance $r_p<100$ $h^{-1}$ kpc,
  the best-fits values
  $\sigma_{\Delta v}=351\pm48$ and $216\pm103$ km s$^{-1}$ 
  for early- and late-type target galaxies, respectively.
It indicates that the velocity limit used for selecting the nearest neighbor
  is large enough not to miss the neighbor galaxies.

The spectroscopic completeness can affect the identification of the genuine nearest neighbor.
The completeness that depends on the apparent magnitude and color as shown in Figure \ref{fig-comp},
  can also depend on the distance between galaxies due to the difficulty in observing galaxies
  close to each other using multi-object spectrograph (MOS).
We checked the completeness as a function of the projected distance to the target galaxy,
  and found that it does not change with the projected distance.
It might be because we combined spectroscopic data from numerous references,
  therefore, the difficulty in observing nearby galaxies using MOS is significantly reduced. 


The virial radius of a galaxy
  is defined as the radius within which the mean mass density
  is 200 times the critical density of the universe ($\rho_c$),
  and is given by
\begin{equation}
r_{\rm vir} = (3 \gamma L  / 4\pi / 200{\rho}_c )^{1/3},
\label{eq-vir}
\end{equation}

\noindent where $L$ is the galaxy luminosity, and
 $\gamma$ is the mass-to-light ratio.
We assume that the mass-to-light ratio of early-type galaxies
  is on average twice as large as that of late-type galaxies
  at the same absolute magnitude $M_B$,
  which means $\gamma$(early)$=2\gamma$(late) 
  [see \S 2.5 of \citet{pc09} and \S 2 of \citet{park08}]. 
The critical density of the universe ${\rho}_c$ 
  is a function of redshift $z$ and
 $\Omega_m(z)= \rho_c (z)/\rho_b(z)=\rho_c(z)/\overline{\rho}(1+z)^3$,
 where $\rho_b$ and $\overline{\rho}$ are the mean matter densities in proper and comoving spaces, respectively.
Then, the virial radius of a galaxy at redshift $z$ in proper space can be rewritten by
\begin{equation}
r_{\rm vir} (z) = [3 \gamma L \Omega_{m,0} / 800\pi \overline{\rho} / \{ \Omega_{m,0}(1+z)^3 + \Omega_{\Lambda,0} \} ]^{1/3}.
\label{eq-vir2}
\end{equation}

\noindent $\Omega_{m,0}$ and $\Omega_{\Lambda,0}$ are the density parameters
  at the present epoch.
We compute the mean mass density $\overline{\rho}$ using
  the galaxies at $z=0.4-0.7$ with
  various absolute magnitude limits varying from $M_B= -17.5$ to $-20.0$.
We found that the mean mass density
  appears to converge when the magnitude cut is fainter than $M_B=-18.0$,
  which means that the contribution of faint galaxies
  is not significant because of their small masses.
In this calculation,
 we weigh each galaxy by an inverse of completeness
  according to its apparent magnitude and color (see Fig. \ref{fig-comp}).
We obtain $\overline{\rho} = 0.0145 (\gamma L)_{-20}$ ($h^{-1}$Mpc)$^{-3}$
  where $(\gamma L)_{-20}$ is the mass of a late-type galaxy with $M_B=-20$.
This final value we adopted is computed using 
  all galaxies with $i_{775W,BEST}\leq23.5$ and $0.4\leq z\leq0.7$
  to account for the contribution of faint galaxies as much as possible.


\begin{figure}
\plotone{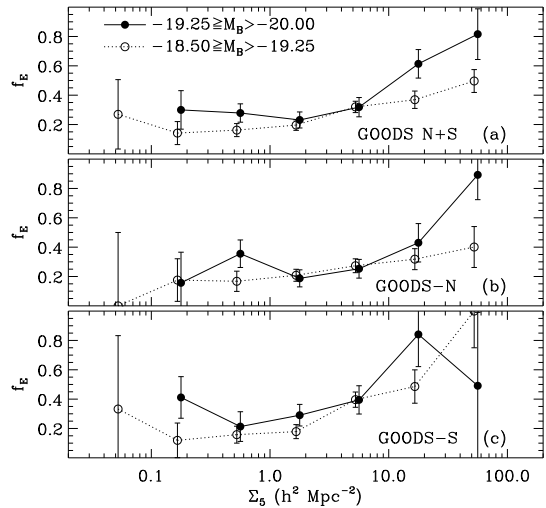}
\centering
\caption{Fraction of early-type galaxies in fixed absolute magnitude ranges 
  as a function of $\Sigma_5$.
  Shown are ({\it a}) GOODS-North plus South, ({\it b}) GOODS-North, and
  ({\it c}) GOODS-South samples.
}\label{fig-fraclarge}
\end{figure}

If we adopt $\Omega_{m,0}=0.27$ and $\Omega_{\Lambda,0}=0.73$, 
  the virial radii of galaxies with $M_B=-18.5$ and $-20.0$ at $z=0$
  are $220$ and 350 $h^{-1}$ kpc for early types, 
  and $180$ and 280 $h^{-1}$ kpc for late types, respectively.
The proper-space virial radii of galaxies with the same luminosities as above, but 
  located at $z=1$, are
  $160$ and 250 $h^{-1}$ kpc for early types, 
  and $120$ and 200 $h^{-1}$ kpc for late types, respectively.

\section{Results}\label{results}
\subsection {Background Density Dependence of Galaxy Morphology}\label{depend}

In Figure \ref{fig-fraclarge}, we plot the fraction of early-type galaxies
  as a function of the background density, $\Sigma_5$.
To account for the incompleteness shown in Figure \ref{fig-comp},
  the early-type fraction is computed by weighing each galaxy by the inverse of completeness
  corresponding to its apparent magnitude and color.
Each of the volume-limited samples is divided into brighter ($-19.25\geq M_B>-20.00$) and
  fainter ($-18.50\geq M_B>-19.25$) subsamples.
The solid and dotted lines are early type-fractions ($f_E$) for
  brighter and fainter subsamples, respectively.
The uncertainties of the fraction represent $68\%$ $(1\sigma)$ confidence intervals 
  that are determined by the bootstrap resampling method.
Figure \ref{fig-fraclarge} clearly shows that 
  the early-type fraction increases along with $\Sigma_5$ (i.e., MDR),
  which is already known in similar high redshift surveys (e.g., \citealt{cap07}).
We note that the overall early-type fraction is higher
  for the brighter subsamples in all surveys.
It implies that the morphology-luminosity relation is already well-established
  at the redshifts under study.
The background density dependence of morphology seems stronger for the brighter galaxies.
In particular, the early-type fraction of the brighter sample rises sharply at high densities
 of $\Sigma_5\gtrsim10$ $h^{2}$ Mpc$^{-2}$.

\begin{figure}
\plotone{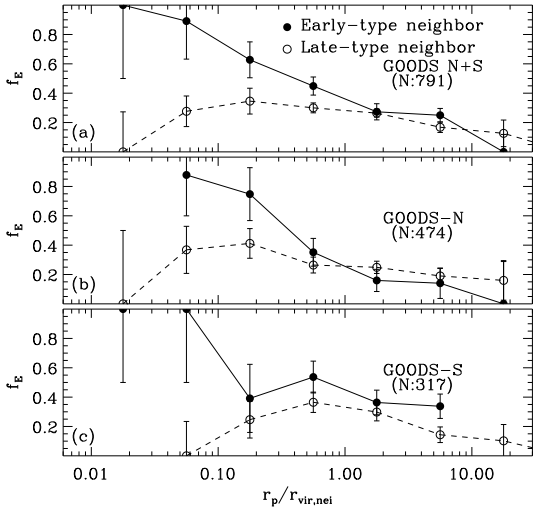}
\centering
\caption{Early-type fraction as a function of the distance to the nearest neighbor galaxy.
The distance is normalized with respect to the virial radius of the nearest neighbor.
Galaxies in the combined GOODS-North plus South sample are used in ({\it a}), 
  those of GOODS-North sample in ({\it b}), and those of GOODS-South in ({\it c}).
}\label{fig-fracneib}
\end{figure}

\subsection {Effects of the Nearest Neighbor}\label{mortrans}

To measure the effects of the nearest neighbor galaxy on the galaxy morphology,
  we plot, in Figure \ref{fig-fracneib}, the fraction of early-type galaxies as a function of
  the distance to the nearest neighbor.
We can see the probability that a galaxy is found to be an early type,
  strongly depends on the projected distance to the nearest neighbor galaxy ($r_p$)
  as well as neighbor's morphology.
When a galaxy is located farther than the virial radius
  from its nearest neighbor galaxy ($r_p\gtrsim r_{\rm vir,nei}$),
  the early-type fraction slowly increases as the distance to the neighbor decreases,
  but its dependence on neighbor's morphology is weak.

On the other hand, when $r_p \lesssim r_{\rm vir,nei}$,
  the early-type fraction increases as the target galaxy approaches an early-type neighbor, but
  decreases as it approaches a late-type neighbor.
It is important to note that the bifurcation of the early-type fraction,
  occurs at $r_p\sim r_{\rm vir,nei}$.
In the case of the cosmology we adopt the radius $r_p=r_{\rm vir,nei}$
  corresponds to the local mass density due to the neighbor of $\rho_n=740\overline{\rho}$.
The bifurcation of the early-type fraction at $r_p\sim r_{\rm vir,nei}$
  is similarly found by \citet{park08} using nearby ($z<0.1$) SDSS galaxies.

\begin{figure}
%
\center
\includegraphics[scale=0.55]{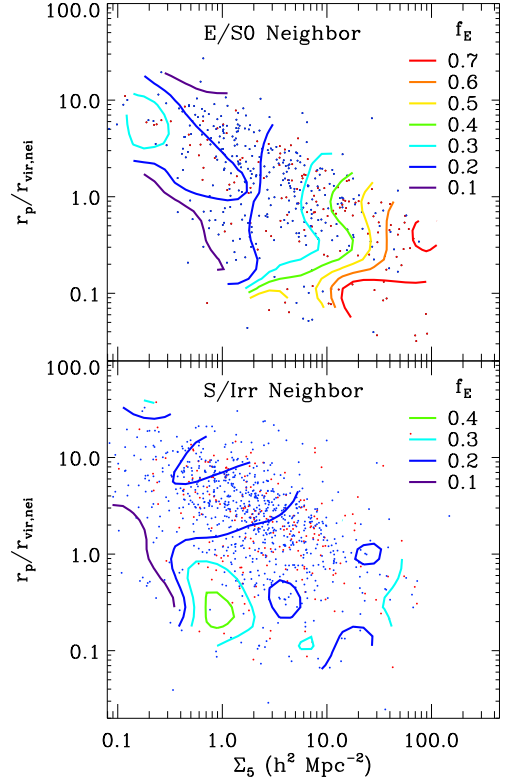}
\caption{
Morphology-environment relation when the nearest neighbor galaxy is 
  an early type ({\it upper}) or a late type ({\it lower}).  Contours show constant early-type 
  galaxy fraction $f_E$. Galaxies used here are brighter than $M_B = -18.5$.
}\label{fig-fracneibden}
\end{figure}

To investigate the effects of large-scale environment on the morphology 
  in company with the effects of the nearest neighbor galaxy,
  we study the early-type fraction in the two-dimensional environmental
  parameter space as shown in Figure \ref{fig-fracneibden}. 
In this case the combined sample of
  GOODS-North plus South is used.
Galaxies are distributed along the diagonal in this figure due
  to the statistical correlation between $r_p$ and $\Sigma_5$. But there is a significant
  dispersion in $r_p$ at fixed $\Sigma_5$.
It is noted that galaxies are located in wide ranges of
  $\Sigma_5$ and $r_p$ for both early- and late-type neighbor cases even though
  the early types tend to have relatively larger $\Sigma_5$ and smaller $r_p$.
In this figure one can study the dependence of $f_E$ on $r_p$ at each fixed value
  of $\Sigma_5$ by moving along a vertical line.

Figure \ref{fig-fracneibden} clearly shows that, when $r_p \la r_{\rm vir,nei}$, 
  the dependence of morphology on the 
  background density and the nearest neighbor distance becomes completely
  different when the neighbor's morphology changes.
When the neighbor is a late type, $f_E$ of target galaxies is about 0.2 and is 
  almost independent of $r_p$ or $\Sigma_5$. But when the neighbor is an early type, 
  $f_E$ can rise up to 0.7 when the target galaxy 
enters a region with the background density exceeding $\Sigma_5 \gg 10 h^2$Mpc$^{-2}$.
There is also a discontinous rise in $f_E$ when the galaxy 
approaches the neighbor closer than about $0.3  r_{\rm vir,nei}$, which corresponds
to $50 \sim 100 h^{-1}$kpc.
When $r_p \gg r_{\rm vir,nei}$, galaxy morphology is independent of all environmental
  factors and $f_E$ is about 0.2.
It should be emphasized that the difference between the left and right panels of
  Figure \ref{fig-fracneibden} is manifest only when $r_p$ is less than about $2 r_{\rm vir,nei}$.
As noted above, a direct effect of the existence of the neighbor on $f_E$ is evident 
at the neighbor distance of $r_p \la 0.3  r_{\rm vir,nei}$ in the case of 
early-type nearest neighbor. 
This confirms the net effects of the nearest neighbor on morphology 
 at fixed large-scale environment for the GOODS galaxies.
However, the effects of late-type neighbors 
do not appear as strong as what is seen for the nearby SDSS galaxies \citep{pc09}. 
More data with a higher completeness are needed to confirm it.

\begin{figure*}
\center
\includegraphics[scale=0.55]{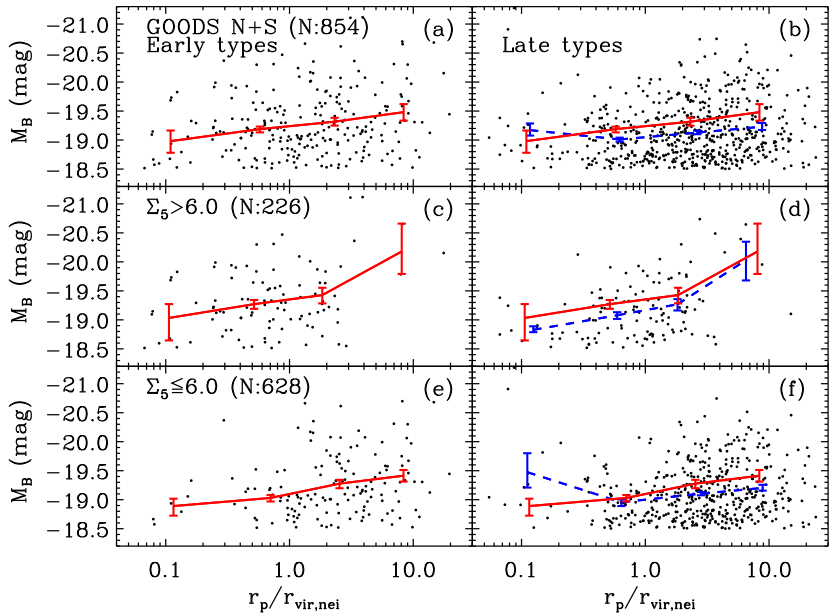}
\caption{Absolute magnitude of early-({\it left}) and 
  late-type ({\it right}) galaxies 
  with $M_B\leq-18.5$ and $0.4\leq z\leq1.0$ in the GOODS-North plus South sample
  as a function of the distance to the nearest neighbor galaxy.
The median magnitudes of early and late types are represented by solid and dashed curves, respectively.
The curves in the left panels are repeated in the right panels as solid curves.
Galaxies in all environments are used in ({\it a}) and ({\it b}), 
  those in high density regions are in ({\it c}) and ({\it d}), and
  those in relatively low density regions are in ({\it e}) and ({\it f}).
Number in parenthesis is the number of galaxies
  in the combined sample or local density subsample.
}\label{fig-lum}
\end{figure*}

  
In Figure \ref{fig-lum}, we plot the evolution-corrected, rest frame $M_B$ as a function of
  the distance to the nearest neighbor for the combined sample of GOODS-North plus South
  with $M_B\leq-18.5$.
The lines show the median value in each projected separation bin.
Top panels show that the early-type galaxies having $r_p>r_{\rm vir,nei}$ are significantly brighter
  than those at $r_p<r_{\rm vir,nei}$,
  while the magnitude difference for late-type galaxies in the two regions is smaller.
If we divide the galaxies into 
  those in relatively higher density region ($\Sigma_5>6$ $h^{2}$ Mpc$^{-2}$)
  and those in lower density region ($\Sigma_5\leq6$ $h^{2}$ Mpc$^{-2}$),
  the increase of galaxy luminosity with increasing $r_p$ is manifest
  in the high density region.

\section{Discussion}\label{discuss}
\subsection{Morphology and Luminosity Transformation}\label{morphdis}

We found in Figure \ref{fig-fracneib} that the morphological types of high redshift galaxies
  depend critically on the small-scale environment,
  which is characterized by the morphology of the nearest neighbor galaxy 
  and the distance to the nearest neighbor galaxy.
If our results are affected by the trend that the early-type fraction ($f_E$)
  is higher when the local density is higher (i.e., MDR),
  $f_E$ should be a function of $r_p$ which is independent of the neighbor morphology,
  and the two curves in the top panel of Figure \ref{fig-fracneib}
  have the same amplitude at each neighbor separation.
The fact that $f_E$ is independent of neighbor morphology at $r_p>r_{\rm vir,nei}$
  but starts to show a significant dependence at $r_p<r_{\rm vir,nei}$,
  demonstrates that the neighbor effects are the dominating factor of 
  the change in $f_E$ and the large-scale background density is not.

Increase of $f_E$ with decreasing the neighbor distance at $r_p\gtrsim r_{\rm vir,nei}$,
  can be explained by the tidal effects of neighbor galaxy.
\citet{park08} showed that tidal energy deposit relative to the binding energy 
  for the dark halos of equal mass galaxies,
  is not negligible at the separation of virial radius.
The tidal effects can accelerate the consumption of cold gas,
  changing late types to early types.
In fact, \citet{pc09} showed that
  the center of late-type galaxies becomes bluer by the existence of neighbor galaxies
  even when $r_p>r_{\rm vir,nei}$ independently of the morphological type of the neighbor.
Their surface brightness and central velocity dispersion increase
  as the neighbor distance decreases, implying the growth of the bulge component.

The bifurcation of $f_E$ at $r_p\sim r_{\rm vir,nei}$ 
  was interpreted as due to the hydrodynamic effects of the nearest neighbor \citep{park08}.
If a target galaxy approaches a late-type neighbor within one virial radius of the neighbor,
  the cold gas of the neighbor can flow into the target galaxy and
  the target galaxy tends to become a late type.
On the other hand, 
  if the target galaxy approaches an early-type neighbor within one virial radius of the neighbor,
  the hot gas and the tidal force of the neighbor accelerate the consumption of cold gas
  so that the target galaxy tends to evolve to an early type.

Moreover, it is noted in Figure \ref{fig-fracneibden}, that 
  the galaxies tend to be early types when they have close early-type neighbors even though
  they are in low density regions, and
  the galaxies are likely to be late types when they have close late-type neighbors
  even though they are in high density regions.
The galaxy morphology appears to depend on the large-scale background density.
But it may be the result of the cumulative effects of neighbor interaction
  that is stronger in high density regions \citep{park08}.

We also found in Figure \ref{fig-lum} that isolated galaxies are brighter than less isolated ones.
These results can provide important hints for the evolution of galaxies.
Previously, \citet{park08} obtained results similar to ours using the SDSS data (see their Fig. 6),
  and suggested a unified scenario as follows.
Once the separation of two galaxies becomes small enough, namely
  $r_p\lesssim0.05r_{\rm vir,nei}$ \citep{pc09},
  they undergo a merger and the merger product will be more massive.
The merger product will typically find itself isolated from its neighbors of comparable mass.
As galaxies experience a series of merger events,
  the cold gas will be exhausted due to the merger-induced star formation.
Thus the massive galaxies are likely to be early types.
As a supporting evidence for this scenario,
  they showed that, at fixed background density, post-merger features such as 
  large displacement of the galaxy nucleus from the center, turmoil features,
  and/or very close double cores,
  are more frequently seen in the isolated galaxies compared to the less isolated ones.

We extended our analysis to another multiwavelength survey, AEGIS,
  of which survey area ($\sim$710 arcmin$^2$) is larger than GOODS ($\sim$300 arcmin$^2$).
Using AEGIS data, we measured the early-type fraction of target galaxies
  as a function of the nearest neighbor distance (e.g., Fig. \ref{fig-fracneib}),
  and found that the early-type fraction does not change significantly 
  with the neighbor distance even though it showed dependence on
  the large-scale background density.
Furthermore, it does not show the bifurcation in accordance with neighbor's morphology.
These results are inconsistent with those from GOODS and SDSS samples (e.g., \citealt{park08,pc09}).
The results from AEGIS may have been significantly affected by 
  the low completeness of DEEP2 survey.
Since the completeness of DEEP2 spectroscopic survey is only about 0.5,
  it is expected that the nearest neighbor is seriously misidentified 
  compared to the case using the GOODS or SDSS data.
Our Monte Carlo experiment shows that the fraction of the misidentified nearest neighbor
  reaches about 50$\%$ when the sample completeness is 50$\%$.
We therefore conclude that 
  it is not appropriate to study galaxy interactions using the DEEP2/AEGIS data.

\subsection{Redshift Evolution of Galaxy Morphology}\label{evol}

If galaxies transform their morphology through a series of interactions and mergers,
  it is expected that the high-redshift galaxies, on average, are less massive and richer in cold gas.
Successive consumption of cold gas in galaxies through interaction and merger events
  is likely to transform late types into early types.
And then the early-type fraction at high redshifts 
  is expected to be lower compared to that at low redshifts, and
  the fraction of blue early types among early types to be higher at higher redshifts 
  (\citealt{park08,cap07}; see also \citealt{con05,con08,lotz08,men04,lee06,puz07}).
Moreover, the early-type fraction is expected to evolve differently
  depending on the background density \citep{park08}.
Since the mean separation between galaxies is relatively smaller and
  the merger/interaction rate is higher in high density regions,
  galaxies are expected to show a stronger time evolution
  in morphology as well as luminosity, color, and SFR in high density regions.
On the contrary, in low density regions where the mean separation between galaxies is larger,
  the time evolution of galaxies is expected to be slower 
  because the interactions and mergers among galaxies are less frequent.


\begin{figure}
\plotone{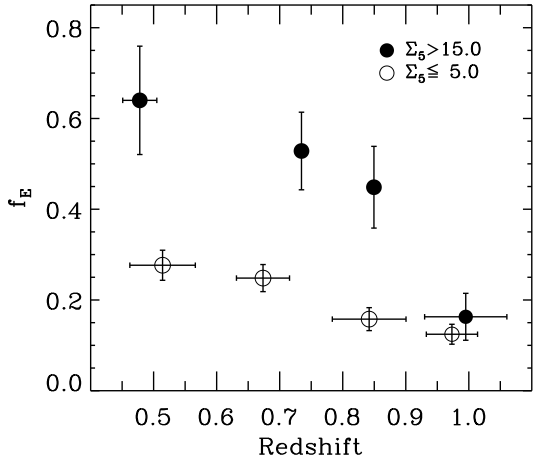}
\centering
\caption{Early-type fraction in the combined GOODS-North plus South sample as a function of redshift.
Filled and open circles indicate the early-type fraction 
  in high and low density regions, respectively, located at the median redshift in each redshift bin.
}\label{fig-fracz}
\end{figure}

There have been several studies of the evolution of morphology
  of high redshift galaxies focusing on the role of environment.
For example, 
\citet{nui05} used S\'ersic index $n$ and the $u^*-g'$ color of galaxies
  as proxies for galaxy morphology to study galaxy evolution.
They found that the number fraction of ``bulge-dominated'' galaxies with $n>2$ in the field
  does not change with redshift, 
  while that in the high density region tends to increase toward the low redshift.
When they use red galaxies with $u^*-g'>1$,
  the fraction of red galaxies increases as redshift decreases in all environments.
However, since they used photometric redshifts to estimate the local density,
  the environment was not accurately determined.
In fact, \citet{coo07} found no time evolution of the fraction of red
  galaxies in low density regions at $0.4<z<1.35$ 
  using spectroscopic redshift data of DEEP2 galaxy redshift survey,
  though the number of red galaxies in their sample may be too small
  to draw a statistically significant conclusion.

Recently, \citet{cap07} used Gini parameter to select early-type galaxies in COSMOS data.
They found that the early-type fraction in all environments grows as redshift decreases,
  and the growth rate is larger in high density regions compared to that in low density regions.
However, in low density regions with $<$100 galaxies Mpc$^{-2}$,
  the early-type fraction at $z>0.4$ changes little with redshift.
\citet{wel07} adopted a combination of S\'ersic index $n$ and bumpiness of galaxies 
  as a proxy of galaxy morphology in SDSS and GOODS-South data.
They found no time evolution of the early-type fraction at $z<0.8$ for both cluster and field regions.
They suggested that the reason why they could not find the evolution of the early-type fraction
  even in high density regions,
  is because they used mass-selected samples and 
  different morphological classification from the other studies.

\begin{figure}
\plotone{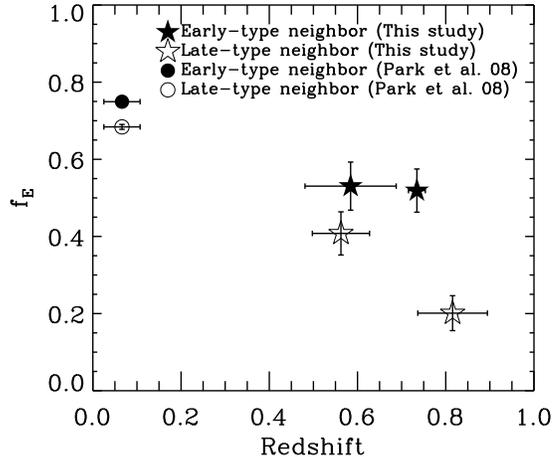}
\centering
\caption{Dependence of the early-type fraction
  on the morphology of the nearest neighbor galaxy
  for the galaxies in the combined GOODS-North plus South sample.
Filled and open star symbols indicate the early-type fraction 
  for the early- and late-type neighbor cases in the GOODS-North plus South sample, respectively.
The redshift bins are $0.4\leq z\leq0.7$ and $0.7< z\leq1.0$, and points are
  at the median redshifts.
Circles denote the early-type fraction for the nearby galaxies in the SDSS \citep{park08}.
}\label{fig-fracneibz}
\end{figure}

To confirm the environmental dependence of evolution of galaxy properties,
  we present the early-type fraction as a function of redshift in Figure \ref{fig-fracz}.
We divide our sample according to the redshift and the background density.
It shows that the early-type fraction in high density regions is higher
  than that in low density regions in all redshift intervals explored.
Moreover, the early-type fraction in high density regions increases
  much more rapidly as redshift decreases.
It indicates a strong evolution of galaxy morphology in high density regions,
  which is consistent with the observational results of some of previous studies
  and also with the predictions of \citet{park08}.

Since our study indicates that the nearest neighbors derive the transformation of galaxy morphology,
  it is important to check galaxy morphology has evolved under strong influence of the nearest neighbors.
Figure \ref{fig-fracneibz} is such an examination.
We divided our sample into two subsets containing galaxies with early- or late-type neighbors.
In this analysis we included only those galaxies that are interacting with neighbors ($r_p<r_{\rm vir,nei}$).
For the early-type fraction at low redshifts,
  we used the SDSS galaxies with $-20.0>M_r>-21.5$ and $z<0.1$
  and whose magnitudes roughly fall into our $M_B$ magnitude range \citep{choi07,park08}.
Figure \ref{fig-fracneibz} clearly shows that
  the early-type fraction monotonically increases as redshift decreases,
  and the early-type fraction for the case of early-type neighbor is always larger
  than that for the late-type neighbor case.
Primordial origin is not likely to be the reason for this systematic difference
  because the neighbor effects are limited within relatively tiny volume of the universe
  associated with each galaxy (see Fig. \ref{fig-fracneib}).
As a comparison, the RMS displacement of matter in the flat $\Lambda$CDM universe is 7.7$h^{-1}$Mpc by z=0.5,
  more than an order magnitude larger than the virial radius of typical galaxies \citep{pk09}.
This result demonstrates an important role of the nearest neighbor 
  in the evolution of galaxy morphology, and
  is consistent with the prediction that 
  the early-type fraction from high to low redshifts will increase
  because a series of interactions and mergers
  tend to transform late types into early types.


In summary,
  the early-type fraction in high density regions increases significantly as redshift decreases,
  while that in low density regions increases much more gently.
The evolution of galaxy morphology is also found to depend critically on the small-scale environment,
  which is characterized by the morphology of the nearest neighbor galaxy 
  and the distance to the nearest neighbor galaxy,
  in addition to the large-scale background density.
A large sample of high redshift galaxies is needed to separate between the roles of
  small- and large-scale environments.
We also found that isolated galaxies are brighter than less isolated ones.
All these results are consistent with the predictions of \citet{park08},
  and confirm their unified scenario of transformation of galaxy morphology 
  and luminosity class at high redshifts.
It implies that galaxy-galaxy interactions play an important role 
  in the evolution of morphology and luminosity classes of galaxies over a long period of time.

\section{Conclusions}\label{sum}

Using the spectroscopic sample of galaxies in the GOODS,
we presented evidence for morphology and luminosity transformation
  of galaxies at $0.4\leq z\leq1.0$.
We determined the morphological types of all high redshift galaxies 
  by visual inspection, and
  used spectroscopic redshifts of galaxies to determine the environmental parameters.
We examined the effects of the nearest neighbor galaxy and the local galaxy number density 
  on the galaxy morphology.
Our main results are as follows:

\begin{enumerate}

\item The early-type fraction increases with 
  the surface galaxy number density estimated from 5th-nearest neighbor galaxies ($\Sigma_5$).
  This confirms the MDR followed by high redshift galaxies ($0.4\leq z\leq1.0$).

\item When a galaxy is located farther than the virial radius
  from its nearest neighbor galaxy,
  the probability for the galaxy to be an early type ($f_E$)
  decreases with increasing distance, and is independent of 
  morphological type of the nearest neighbor.

\item When the separation with the nearest neighbor galaxy is smaller than the virial radius 
  of the neighbor,
  $f_E$ increases as the target galaxy approaches an early-type neighbor, but
  tends to stay constant as it approaches a late-type neighbor.
Conformity in morphology between neighboring galaxies is confirmed at high redshifts.
The realm of conformity is confined within the virialzed region associated with
  each galaxy plus dark halo system.

\item We find that more isolated galaxies are more luminous.
It can be explained by the luminosity evolution of galaxies through a series of mergers.

\item The early-type fraction $f_E$ increases very rapidly as redshift decreases
  in high density regions, but increases only mildly in low density regions.
At $z>1$ the MDR seems very weak, and it should be confirmed by higher redshift data.
Our findings are consistent with the prediction of \citet{park08} that 
  the early-type fraction evolves much faster
  in high density regions than in low density regions because
  the rate of galaxy-galaxy interactions is higher in high density regions
  and a series of interactions and mergers through a cosmic time
  transform late types into early types.

\end{enumerate}

The critical role of galaxy-galaxy interactions in the evolution of
  galaxy properties has been confirmed in the general environment \citep{park08,pc09}
  and in the galaxy cluster environment \citep{ph08} at low redshifts.
The results of the present work extend the previous findings to high redshifts.
We plan to further extend our analysis to even higher redshift universe 
and to high redshift clusters.


\acknowledgments
We thank the anonymous referee for constructive comments that 
  helped us to improve the manuscript.
CBP acknowledges the support of the Korea Science and Engineering
Foundation (KOSEF) through the Astrophysical Research Center for the
Structure and Evolution of the Cosmos (ARCSEC).


\end{document}